\documentclass{article}

\usepackage{arxiv}

\usepackage[utf8]{inputenc} 
\usepackage[T1]{fontenc}    
\usepackage{hyperref}       
\usepackage{url}            
\usepackage{booktabs}       
\usepackage{amsfonts}       
\usepackage{nicefrac}       
\usepackage{microtype}      
\usepackage{lipsum}
\usepackage{graphicx}

\title{A Novel IoT Sensor Authentication Using HaLo Extraction Method and Memory Chip Variability}

\author{
  Holden~Gordon \\
  Department of Electrical and Computer Engineering\\
  Santa Clara University\\
  Santa Clara, CA 95050 \\
  \texttt{hgordon@alumni.scu.edu} \\
    \And
  Thomas~Lyp\\
   Department of Electrical and Computer Engineering\\
  Santa Clara University\\
  Santa Clara, CA 95050 \\
  \texttt{tlyp@scu.edu} \\
  \And
  Calvin~Kimbro\\
   Department of Electrical and Computer Engineering\\
  Santa Clara University\\
  Santa Clara, CA 95050 \\
  \texttt{ckimbro@slumni.scu.edu} \\
  \And
  Sara~Tehranipoor\\
   Department of Electrical and Computer Engineering\\
  Santa Clara University\\
  Santa Clara, CA 95050 \\
  \texttt{ftehranipoor@scu.edu} \\
}

\begin{document}
\maketitle

\begin{abstract}
	
  In this paper, we propose flash-based hardware security primitives as a viable solution to meet the security challenges of the IoT and specifically telehealth markets. We have created a novel solution, called the High and Low (HaLo) method, that generates physical unclonable function (PUF) signatures based on process variations within flash memory in order to uniquely identify and authenticate remote sensors. The HaLo method consumes $60\%$ less power than conventional authentication schemes, has an average latency of only 39ms for signature generation, and can be readily implemented through firmware on ONFI 2.2 compliant off-the-shelf NAND flash memory chips. The HaLo method generates 512 bit signatures with an average error rate of $5.9 * 10^{-4}$, while also adapting for flash chip aging. Due to its low latency, low error rate, and high power efficiency, the HaLo method could help progress the field of remote patient monitoring by accurately and efficiently authenticating remote health sensors.
  
\end{abstract}

\section{Introduction}
\label{intro}
Internet of Things (IoT) sensors has seen explosive growth over the last twenty years. The consumer IoT market is estimated to reach 142 billion dollars by 2026 at a CAGR of $17\%$. Estimates forecast that by 2025, there will be 152,200 IoT devices connecting to the internet per minute. However, the expanding IoT landscape has similarly created an increased attack surface. Malware attacks targeting IoT have increased by $30\%$ with $76\%$ of IoT risk professionals believing their organization’s IoT security posture leaves them vulnerable to cyber attacks~\cite{puf-sec}. This is particularly important for the subset of IoT devices such as medical sensors that are used in telehealth applications~\cite{hassan2017internet, tehranipoor2017exploring, wortman2017proposing, tehranipoor2017investigation}.

Telecommunication in healthcare, or telehealth, provides a means by which patients can interact with medical professionals and health-related services virtually. In light of the COVID-19 worldwide pandemic, the need for secure and readily available remote patient monitoring has never been more important. Rural and low-income communities, in particular, have been severely impacted by the lack of accessibility to in-person healthcare. This has created the need for access to remote patient monitoring and virtual health visits in order for greater accessibility to premier care. However, the convenience of connecting medical providers with patients remotely also introduces significant security and privacy risks. One such risk of using unsecured medical devices is the potential for a major privacy breach, as they store sensitive information such as vital signals, diagnosed conditions, therapies, and a variety of personal data~\cite{TelehealthII}. For example, on the dark web, an individual's private health information (PHI) goes for 20 times to 100 times the value of a social security number or credit card. This creates a strong financial incentive for malicious actors to illicitly steal healthcare data from small embedded telehealth sensors which are the most exposed computing elements due to the tight processing, energy, and latency requirements of IoT devices~\cite{anagnostopoulos2018securing}. 

Physical Unclonable Functions (PUFs) can provide a lightweight and tamper-proof security primitive for IoT devices particularly telehealth sensors~\cite{puf-sec, tehranipoor2018towards, tehranipoor2017design11}. PUFs create cryptographic signatures that can be used for authentication~\cite{yan2015novel}, software attestation, and cryptographic key generation~\cite{DRAM-PUF, aguirre2020systematic}. These signatures are derived from the sub-micron process variations present in integrated circuits (ICs) ~\cite{DLA-PUF}. These signatures are never stored on the device itself and in many cases are extremely difficult to tamper~\cite{logic-locking,DRAM}. 

In this paper, we propose a novel PUF extraction scheme called the High-Low method (HaLo), which extracts process variations found in commercially available NAND flash memory chips. The HaLo method aims at minimizing authentication latency and maximizing the lifetime of the flash chip by limiting program/erase cycles. The HaLo method is also proven to work using commercially available NAND flash chips while being entirely controlled by a microcontroller that costs under \$15, making it one of the most cost effective published flash-based solutions to our knowledge. Our main contributions are summarized as follows:

\begin{itemize}
\medskip
  \item We have built a novel PUF extraction technique called the HaLo method, which supports edge deployment on low-cost microcontrollers. This will lower the cost of entry, and help encourage secure data transmission for remote devices.
  \item The proposed HaLo method offers a PUF solution for accurate authentication and has lower latency and lower power consumption than other PUF generation techniques.
  \item The HaLo method is compatible with off-the-shelf ONFI 2.2 compliant flash chips, which could lead to backward compatibility implementation on existing health sensors.
  \end{itemize}
  
\medskip
The organization of the remainder of this paper is as follows. First, \textbf{Section II} will discuss the preliminary background information required in order to understand the HaLo method. Next, \textbf{Section III} will give an in-depth look at related published authentication schemes using process variation, as well as a look at the advantages of the novel HaLo extraction method. \textbf{Section IV} will discuss the HaLo extraction method in detail including design constraints and considerations, and \textbf{Section V} will provide experimental results and validation of the novel HaLo method. Finally, \textbf{Section VI} will provide details on the telehealth application proof of concept built with the HaLo extraction method, and \textbf{Section VII} will give a brief conclusion and summary of the work, along with details on proposed future work for this project.

\section{Preliminaries}
In this section, we will briefly discuss the general background information involving our proposed authentication scheme. Specifically, this section covers an introduction to current wireless authentication protocol solutions, a brief understanding of process variations found in flash memory chips, and a general understanding of PUFs~\cite{gordon2021flash}.

\subsection{Wireless Authentication Methods}
There are a variety of wireless authentication methods that are used by IoT and Telehealth devices. One of the most common ways to authenticate these resource constrained devices is by utilizing pre-shared keys~\cite{Wifi-PSK-Hack}. Pre-shared keys are authentication keys or tokens that are shared with a device prior to its deployment. These keys are then exchanged with a gateway in order to authenticate an IoT device. There are several important vulnerabilities within this model that are solution wishes to address. First, the keys can be extracted out of firmware by a sophisticated hacker who has access to the physical device. This has happened in the industry with Link Plugs and other smaller IoT devices~\cite{link}. Secondly, these keys can be cloned through replay attacks and deauthentication attacks~\cite{Wifi-PSK-Hack,Adversarial}. This was shown to be effective on all WEP Wifi routers which extracted wifi keys due to the lack of 'freshness' in the messages sent between endpoints and wifi routers. By using PUFs and TRNGs, many of these prior security vulnerabilities can be drastically mitigated by providing random nonces and authentication signatures~\cite{wortman2020exploring}. Firstly, the PUFs allow secrets to be embedded in process variations which are not stored in any nonvolatile memory which prevents hackers from simply dumping onboard firmware and finding pre-shared keys. Secondly, the onboard TRNG mitigates replay attacks by preventing attackers from arbitrarily replaying authentication messages.

\subsection{MLC NAND Flash Memory Architecture}

\begin{figure}[t]
		\centering
		\includegraphics[width=0.56\linewidth]{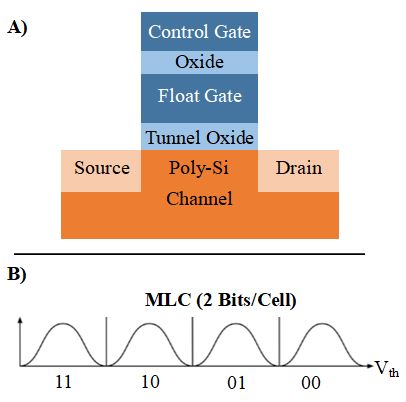}
		\vspace{-7pt}
		\caption{A) Overview of NAND Flash Cell B) MLC Flash Chip Digitization}
		\label{fig:flashcell}
        \vspace{-10pt}
\end{figure}

Flash NAND memory is a type of non-volatile memory that stores user data in the physical form of charge on a \textit{float gate}. Programming these cells requires electrons to move from the polysilicon channel on/off the floating gate via electron tunneling, described in figure~\ref{fig:flashcell}A. It is important to note that this electron tunneling can damage the tunnel oxide, which means that flash cells become less reliable after many program/erase cycles (PECs). Most flash memory is rated anywhere from 1,000 to 10,000 PECs. A flash's lifetime can increase drastically by using a memory management controller that distributes PECs evenly across all cells called wear leveling.

In the case of \textit{Multi-Level Cell (MLC)} flash chips, 2 bits of data are stored on each cell, where the current float gate voltage is compared to multiple threshold voltages in order to determine the cell's digital value, as shown in Figure~\ref{fig:flashcell}B. Flash has high memory density, low cost, and ability to be programmed and erased electronically without moving parts, flash NAND memory has exploded in popularity for remote devices and IoT systems.

\subsection{Process Variations and PUFs for Flash Memory}
Like all other silicon-based ICs, flash memory chips are subject to many different forms of process variations. In general, most process variations are uncontrollable imperfections caused by limitations in modern lithography processes. These variations are commonly caused by various disturbs caused by parasitic capacitance. By performing multiple read or program operations on sections of the flash chips, disturbs can be induced on these chips causing random fluctuations and bit flips. These fluctuations depend on each cell's relative gate thickness and width. both of which are uncontrollable and extremely hard to model. Finally, this allows for unique signatures that can be generated for each page of the flash memory. These signatures are known as PUFs which can be used for secure key generation and authentication mechanisms~\cite{FlashVariation}. Each PUF has a challenge and response. The challenge is the input to the function which then outputs an unclonable response. Key mathematical metrics are used to describe PUF efficacy which will be touched on in more detail in \textbf{Section VI}. 


\subsection{Physical Unclonable Functions (PUFs)}
A PUF is a function that produces a unique signature based on challenge-response pairs. These unique signatures often referred to as \textit{silicon fingerprints}, are by-products of intrinsic, uncontrollable process variations found on a silicon-based IC~\cite{ProcessVariation}. As mentioned in a previous section, these process variations are present on every chip, and they can greatly differ from one another in regards to performance. While understanding specific instances of process variations can help influence the design for PUF extraction, the absolute variations do not typically matter. 

\section{Related Works}
Flash memory has gained popularity in recent years due to its cheap cost and high density for storage applications. This is particularly the case for Flash NAND memory~\cite{DRAM-Overview}. Rather than including extra CMOS as a PUF, utilizing the onboard Flash memory can conserve space and energy for an extremely resource-constrained devices such as those seen in telehealth applications. 

When looking at the development of different Flash PUFs, there are several important trends to recognize. The first Flash PUFs were created from 2012-2017~\cite{Prabhu,Wang,Jia,Saito}. These seminal works were predominately focused on showing how Flash memory was a viable candidate for memory-based PUFs. These Flash PUFs were important for showing the promise of flash memory, but they struggled to account for several factors such as aging, temperature, design constraints. For example, Prabhu et al.~\cite{Prabhu} required hundreds of thousands of programs in order for PUF signatures to develop which can take several hours. Similarly, Kim et al.~\cite{Kim} required very fine-grained flash programming which required knowledge of sensing voltages that are typically proprietary on commercial flash chips. Similarly, Wang et al.~\cite{Wang} created both a novel PUF and TRNG but incurred very high processing overheads that leads to high latency. 

From 2018 to 2020, Flash PUF development began marked by the work of~\cite{Wu,Clark,Poudel,Mahmoodi,Sakib,Larimian}. Many of these proposed designs significantly enhanced Flash PUF architectures that took into account many different factors. These Flash PUFs incorporate advanced parameters such as aging resistance, temperature resistance, and unique architectures. For example, work from Clark et al.~\cite{Clark} designed a Flash PUF that was voltage resistant. Secondly. Poudel et al.~\cite{Poudel} designed a Flash PUF that works on the onboard microcontroller NOR Flash Memory. Similarly, Larimian et al.~\cite{Larimian} verified the machine learning resilience of their Flash-based PUFs by performing extensive deep learning tests. Finally, Mahmoodi et al.~\cite{Mahmoodi} created one of the most stable and resilient Flash PUFs by modifying the cells to extract leakage current.

However, several open challenges remain. In order to make PUFs commercially viable, they must be generated using cheap commercial microcontrollers in order to keep their price point down. Furthermore, it is helpful that these PUFs can be deployed on legacy systems by using commercial off-the-shelf flash memory and these systems should avoid excessively long latency such as those seen in ~\cite{Prabhu}. Although, even some of the most cutting-edge work do not use commercial off-the-shelf components. This can reduce the adoption of Flash PUF technology. Secondly, many of these Flash PUFs require slowly building charge differences on the floating gate through hundred or thousands of program/erase cycles. This is effective at generating signatures; however, this drastically increases the latency required to generate these PUF responses. Finally, many of the systems that avoid the slow build-up of charge have to use expensive FPGAs with Gigahertz clocking speeds to generate fine-grained interrupts as shown in~\cite{Clark}. This is not viable for edge deployments particularly those such as telehealth-based applications.  

Our work seeks to bridge this gap by developing a Flash-based PUF that uses only ONFI standard commands; works on a 15 dollar microcontroller; uses off the shelf commodity flash chips~\cite{yan2020bit}, generate signatures with low latency using a novel interrupt technique; and is able to compete with some of the most resilient PUF structures identified in previous work.

\section{HaLo Flash PUF Extraction Technique}
HaLo extracts process variations found in flash NAND memory cells for authentication applications. This section will discuss the HaLo experimental setup, design considerations for PUF extraction, and the HaLo method itself.

\subsection{Experimental Setup}
As mentioned in our summarization of contributions section, it is vital to keep the experimental setup as minimal as possible in terms of both cost and complexity. While previous works have mostly required high clock frequencies and custom chips, our novel design uses a 100 MHz clock and a simple TSOP Adapter in order to connect to the flash chip. Our simple bill of materials consists of:
\begin{itemize}

  \item STM32 based microcontroller with 100 MHz maximum clocking frequency. This will serve as the memory controller and is used to read/write/erase the flash memory.
  \item 32 GB MLC Flash NAND memory chip from Micron that does not have an integrated memory controller. The memory chip can read/store application data and is also used in the HaLo extraction method
  \item TSOP Adapter in order to interact with packaged flash NAND chip

\end{itemize}

The entire experimental setup is purchased for just shy of \$20, and all items were readily available for purchase to be included in a commercial application. It is worth noting that we used an unmanaged flash chip in order to give us full control of the device. This extra control allowed us to write directly to specific memory locations without worrying about wear-leveling control or error-correcting code (ECC), which are implemented in most managed flash components. This gave us the highest control on interrupting the writing, erasing, and reading processes, because we were directly controlling the chip's behavior without interference from the memory controller. The HaLo method only requires the most basic memory access and modification functions including read page, write page, and erase block. 

\begin{figure}[t]
		\centering
		\includegraphics[width=0.68\linewidth]{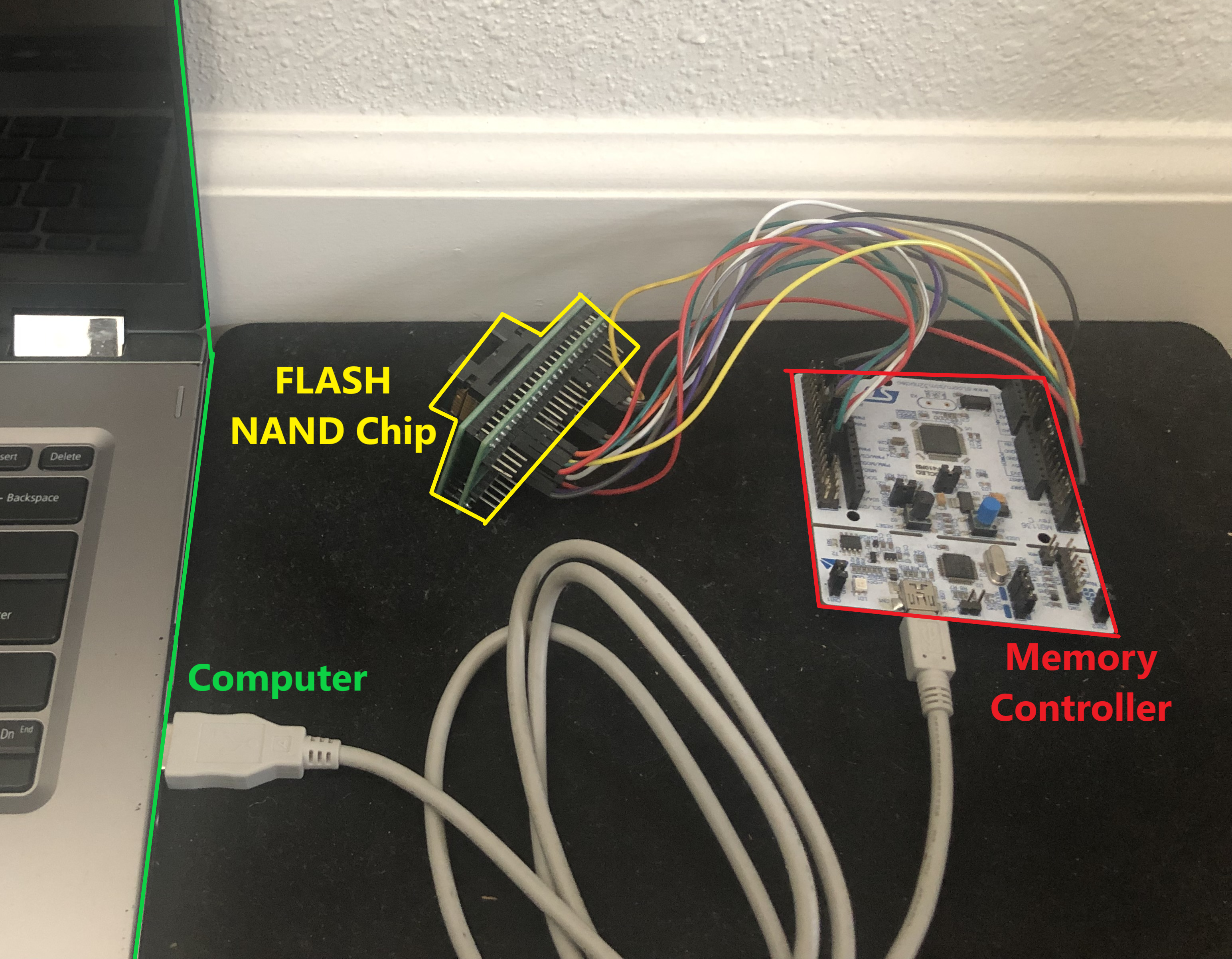}
		\vspace{-7pt}
		\caption{Experimental Setup Diagram}
		\label{fig:setup}
        \vspace{-10pt}
\end{figure}

The experimental setup is shown in Figure~\ref{fig:setup}. The memory chip interface was \textit{Open NAND Flash Interface (ONFI)} version 2.2, so the interrupt sequence and reading of data are ubiquitous across all ONFI 2.2 memory chips. ONFI 2.2 requires 7 control signals and an 8-bit bus for data, as shown with jumper cables connecting the Flash NAND Chip to the Memory Controller. Additionally, a USB connection was made between the Memory Controller and a computer. PUF trial data was transferred via the USB cable so that all statistical testing could be performed on the computer. A computer is simply an analysis tool, and no part of the actual PUF extraction method is performed on the computer.

The flash chip tested in this work stores pages consisting of 4096 bytes of data and 224 extra bytes for ECC. Blocks are organized in chunks of 256 pages, with each 32 Gb chip having access to 4096 unique blocks ~\cite{DRAM-TRNG,COTS-DRAM}. 

\subsection{Proposed PUF Extraction: Design Considerations}
With the experimental setup described in the previous section, work began on the PUF extraction method itself. When designing the PUF extraction method, there are two major metrics that are primarily considered. The first feature was minimizing the number of program/erase cycles (PECs) required in order to extract a reliable signature. As mentioned in the \textit{Preliminaries} section, flash memory devices have a limited lifetime measured in PECs. As the charge is tunneled onto and off of the floating gate, the tunnel oxide that separates the silicon channel and the floating gate begins to deteriorate. As the deterioration continues, the data held within each cell becomes unreliable due to the increase of charge leakage on the floating gate. Wear on the cells is an issue for data retention and general application use, but it also poses an issue with unreliable signature extraction, because the programming and erase times for each cell are altered as the cell reaches its end of life. In order to combat this aging effect, HaLo was designed in order to require as few PECs as possible.

The second major consideration was lowering the signature extraction latency. Faster signature extractions allow for faster authentication and lower power consumption ~\cite{RO}. Telehealth sensors in the scope of this work can be treated as edge devices that are resource-constrained by both processor speed as well as limited battery life. In order to prolong the sensor's battery lifetime as well as create a reasonably fast authenticated connection, the method needs to be lightweight and fast.

While additional results and metrics will be discussed in detail in the \textit{Experimental Results and Validation} section below, these two design considerations helped guide the construction of the HaLo method. In the next section, the HaLo technique will be explained in detail, and additional design constraints and considerations will be discussed.

\subsection{PUF Extraction Observations and Techniques}
\begin{figure}
\center
\includegraphics[width=0.5\linewidth]{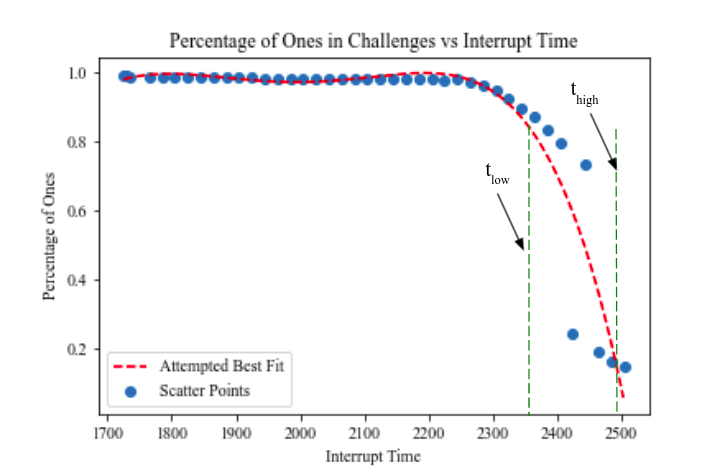}
\caption{Ratio of 1's and 0's for Different Program Interrupt Times}
\label{clocking_granularity}
\end{figure}

As mentioned in the \textit{Related Works} section, the PUF extraction must have low latency and can only use standard edge deployable microcontrollers within the 100MHz frequency range. This introduces several important design challenges in order to make Flash PUFs achievable on low-cost microcontrollers. First, the low latency requirement prevents the design from utilizing hundreds of repeated program and read cycles that slowly increase the charge on the floating gloat. Therefore, our scheme must use fine-grained interrupts in order to interrupt the operation of the cells to force the flash into unsteady positions. However, a single interrupt scheme such as the one seen in Clark et al.~\cite{Clark} does not have a high enough clock resolution to interrupt the flash programming fast enough to generate a 50/50 split between ones and zeros. In fact, signatures that are generated from a single programmed interrupt are either about $80\%$ 1's or $80\%$ 0's. This is highlighted in Figure~\ref{clocking_granularity} where the sysTick clock on the microcontroller is interrupted at different times. As shown in the figure, the clock does not have the granularity to interrupt the programming operation on the microcontroller to generate an even distribution of 1's and 0's. This creates two distinct regions detonated by a low programming interrupt and a high interrupt programming interrupt where the low interrupt generates signatures with about $80\%$ ones and the high programming interrupt generates signals with about $80\%$ zeros. The steep drop-off occurs within a single sysTick clock instruction. This makes interrupting the program a distinct challenge.

Secondly, the lack of granularity has another limiting effect as well. The signatures generated from a single interrupt are extremely noisy. This is due to the interrupting method. This error rate can approach $10\%$ within 80 total P/E cycles. This is due to the lack of granularity in the clock itself. Due to the max 100 MHz clocking speed, an interrupt at a particular clocking delay can vary due to the limited frequency. This causes unintended errors for cells that are sensitive to the interrupt. This necessitates several programs in the enrollment to ensure that the bytes are stable across a small variation in the actual interrupt delay.

Although the error increases rapidly and the clocking speed cannot generate signatures with $50\%$ ones and zeros, several more insights and solutions were generated to remedy these design challenges. Specifically, a well-defined enrollment scheme is designed to select only the most stable cells which can be decoded as one or zero. These can then be used for highly stably signatures. 
 
 \subsection{Proposed PUF Extraction: Enrollment Scheme}
\begin{figure}
\center
\includegraphics[width=0.6\linewidth]{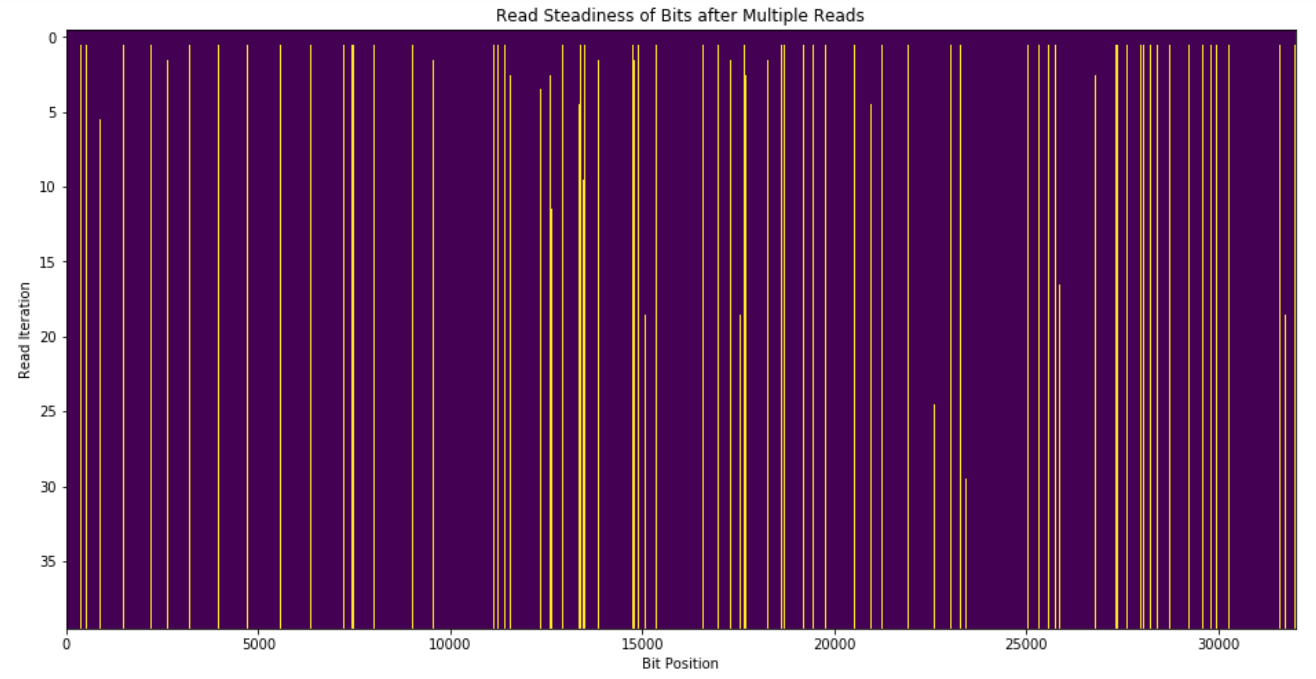}
\caption{Error Increase Over 100 Trials}
\label{stability_read}
\end{figure}

 As other work has shown in Poudel et al.~\cite{Poudel} unstable cells can be identified for TRNG bits by applying several reads. These reads apply a smaller voltage to the floating gate of the flash cells which only slightly disturbs the cells. This can quickly identify unstable cells that are flagged during enrollment. Furthermore, approximately $95\%$ of these unstable bits flip within five reads. Therefore, only applying five reads is sufficient for identifying stable cells through successive read operations. Figure~\ref{stability_read} highlights this observation. This figure graphically shows all 32000 bits on a single page. All of the yellow lines indicate a flipped bit. Many of these bits flip within the first five reads which is an effective filtering process for identifying stable bits within a single read. Combining multiple reads can help identify unstable bits per page along with multiple programs which helps to identify unstable bits due to lack of clocking granularity.
 
 Secondly, flash cell failure is highly spatially dependent. This is evidenced by plotting a histogram of the distance between each cell flip or failure. Failures tend to cluster in groups. This can be modeled as a negative exponential distribution between errors. Therefore, instead of flagging individual bits that are stable, bytes are flagged as stable and only if a byte is completely stable is it passed through the enrollment process. 
 
Combining these observations, the first enrollment strategy was crafted. An interrupt program is used on the low side and another is applied to the high side. Then five reads are performed. If bytes remain stable, across the low interrupt and the high side interrupt they are flagged as stable. After this technique is applied about 1500 bytes are extracted per page which has a total of 8000 bytes. Approximately 800 of these bytes are over $98\%$ accurate during testing. However, approximately 700 bytes are underneath this threshold. This is due to the fact that many stable bytes are incorrectly flagged during the enrollment process. However, a slight wrinkle to the first algorithm is able to generate highly stable PUF responses. This leads to our new enrollment strategy called the HaLo method. 
 
\begin{figure}
\center
\includegraphics[width=0.5\linewidth]{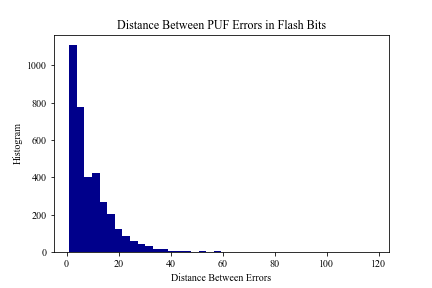}
\caption{Histogram of Distance Between Sequential Failures}
\label{poisson-dist}
\end{figure}
 
\begin{figure}
\center
\includegraphics[width=0.65\linewidth]{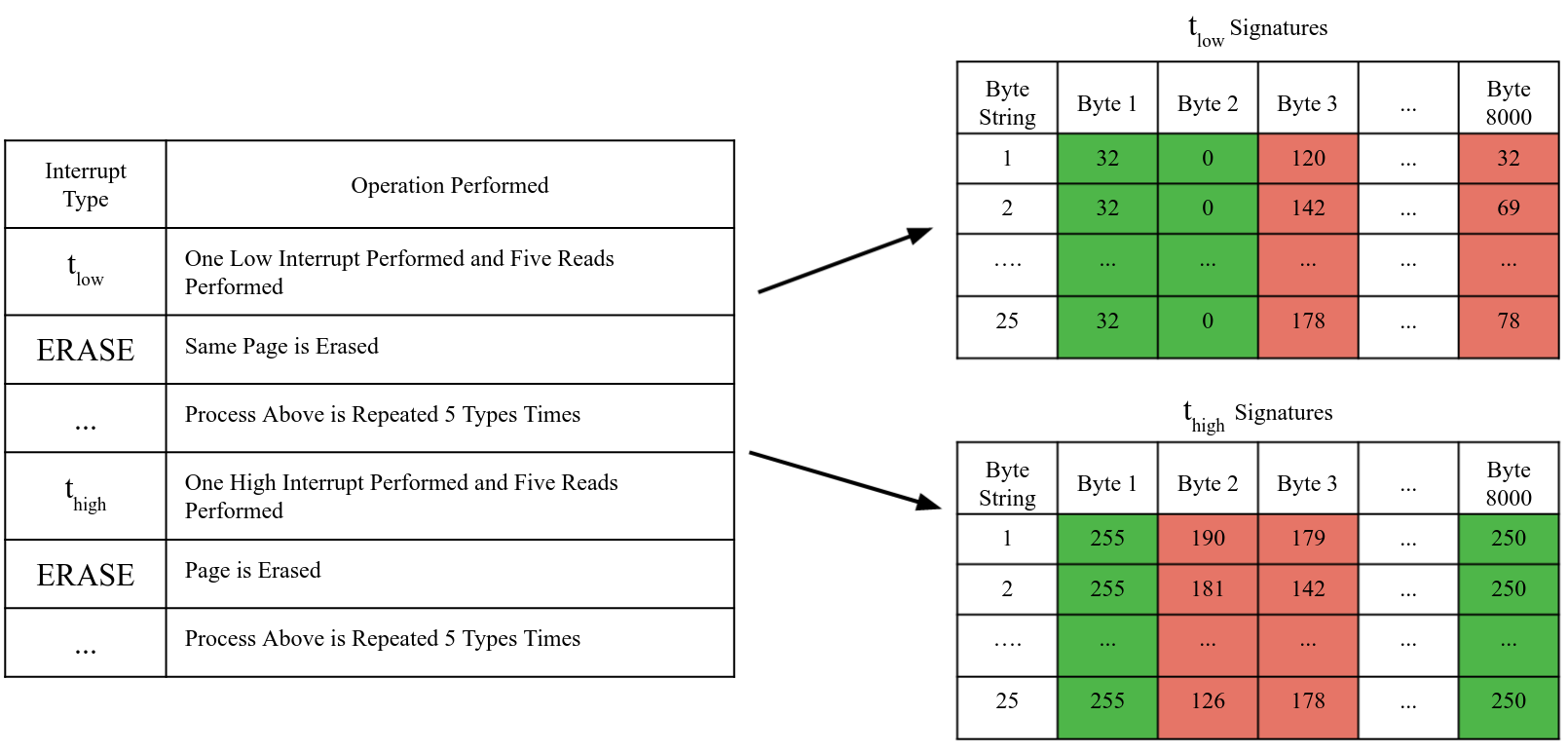}
\caption{This figure highlights the enrollment process where low interrupt and high interrupt signatures are generated to provide a 'map' of stable low byte and high byte values. These values are set along with the page number to compromise a challenge. }
\label{enroll}
\end{figure}

The HaLo extraction method uses a novel technique to generate highly reliable signatures that begins with a byte selection process. Reiterating the previous strategy, rather than aborting the program early and using signatures with $80\%$ ones, two different types of interrupted programs are performed per page: the high interrupt and low interrupt. The low program generates a signature with $80\%$ ones and the high program has signatures with approximately $20\%$. This is repeated five times on each low programming interrupt and high programming interrupt. Since errors tend to be close together, only bytes that are entirely stable are chosen. This creates two separate sets each of which contains 25 signatures due to the five reads per five programs. Next, the stable high bytes locations and the stable low byte locations for a single page are identified. Then bytes that are either highly resistant to programming or highly susceptible to over-programming are chosen. This information is captured by selecting bytes that fully resist programming in the under-programmed section also known as the low side and which bytes are easily programmed in the over-programmed section also known as the high side. If a byte group is simultaneously fully programmed in the over-programmed section and is fully under-programmed in the under-programmed section it is also removed during the enrollment process. This is because greater than $97\%$ of errors are from the aforementioned 700 bytes which are stable in both sections. When referencing Figure ~\ref{enroll}, the only byte selected from the low side is Byte 2 and the Byte selected for the high side is Byte 8000. This then creates a 'map' for each page of the stable low bytes and stable high bytes. 
 
\begin{figure}
\center
\includegraphics[width=0.65\linewidth]{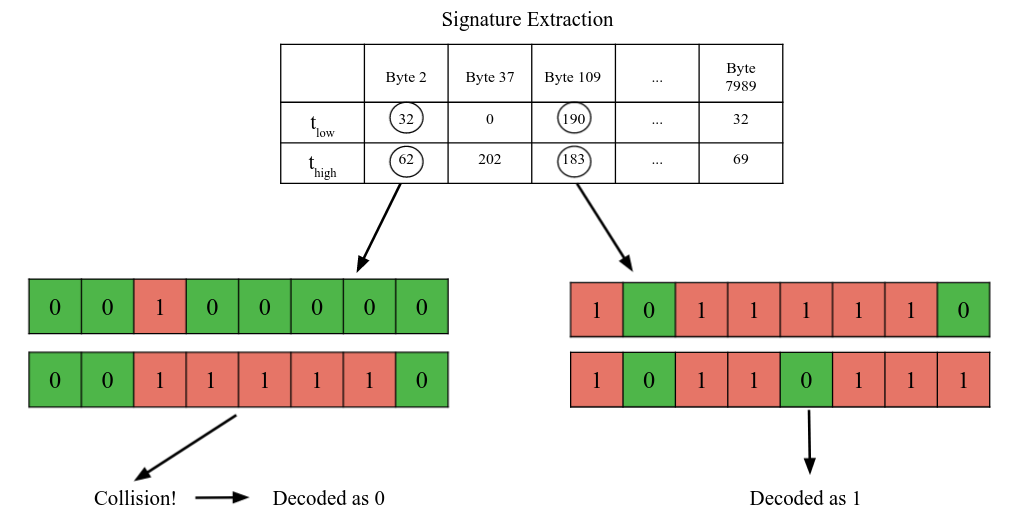}
\caption{This figure highlights the decoding process for a device generating a response to a PUF challenge. First, the device applies the two programming intervals to the page requested in the challenge. Then, the stable bytes are located and are also sent in the challenge to the end device. The device then applies the programming and chooses the majority represented bit in each signature. In the event of a collision the byte that has the strongest weight is selected. This is why the collision scenario in the figure decodes to 0. }
\label{decode}
\end{figure}

 After enrollment, the telehealth device receives a list of byte locations to provide a high program and low program along with the page number to apply the programming. It will do two separate programs: one high and the other low. Then, the majority bit is selected from each byte and is decoded properly. Collisions are minimized since 'mapped' byte values are either highly resistant to over-partial programming or susceptible to it. However, if it occurs the byte with the most amount of parity bits when decoded will be selected. Furthermore, any undefined states (such as equal hamming weights of four) are decoded as low since more errors come from the high side than the low side. This process is highlighted in Figure ~\ref{decode}. This process reduces the percent error by several magnitudes of 10 to around $10^{-4}$. A major downside to this extraction process is that it consumes a tremendous amount of bits to extract only the most stable bits and performs two programs instead of one. Specifically, each page consists of 32,000 bits has an output response of about 500 bits. 

\subsection{Proposed PUF Extraction: Challenge and Response}

After the HaLo enrollment process is complete a 'map' of the stable bytes is stored on the gateway. Each byte in the map is either extremely susceptible to over-programming or highly resistant to it. With this information, the gateway will request particular byte locations from the sensor for authentication. Approximately 256 low byte locations and 256 high byte locations will be sent along with the specific page number that is used. This compromises approximately 600 bytes of space which can be sent in the payload of a single Ethernet frame to the sensor. The microcontroller will then apply two programs and ten total reads and identifies which bytes are stable and which are not. If a byte location produces a majority of ones it is subsequently decoded as one and is assumed to be one of the high side bytes. Inversely, if a byte produces a majority of zeros than it is assumed to be from the low side and is decoded as zero. This extraction technique can resist a maximum of three errors before a byte is incorrectly decoded. Therefore, the gateway will receive a bit string of ones and zeros to the sensor. 

This design allows for two important features for the gateway. Firstly, the gateway can control how long of an authentication response it needs. This can allow our application to adapt to various levels of required security. For example, certain cryptographic applications may only require 100-bit signatures. The gateway then only has to send 100 bytes to the sensor. On the other hand, sensitive security tasks that may demand 512-bit signatures can also be accomplished. Secondly, many other approaches use helper data such as Hamming Codes, Fire Codes, or more recently Low-Density Parity Codes, to recover any errors sent back from the device. These codes leak polarity information about the response values which allows for error correction. Our scheme leaves out any polarity data and simply sends bytes to read. This makes our helper data significantly more robust to side-channel or modeling attacks since the helper data never reveals any polarity information from the bytes it is requesting. However, it is important to note that this advantage is realized because the HaLo enrollment algorithm filters bytes from pages very aggressively. On average, each page returns approximately 700 usable bytes. Therefore, about $83\%$ bytes are not usable.

\begin{figure}
\center
\includegraphics[width=0.6\linewidth]{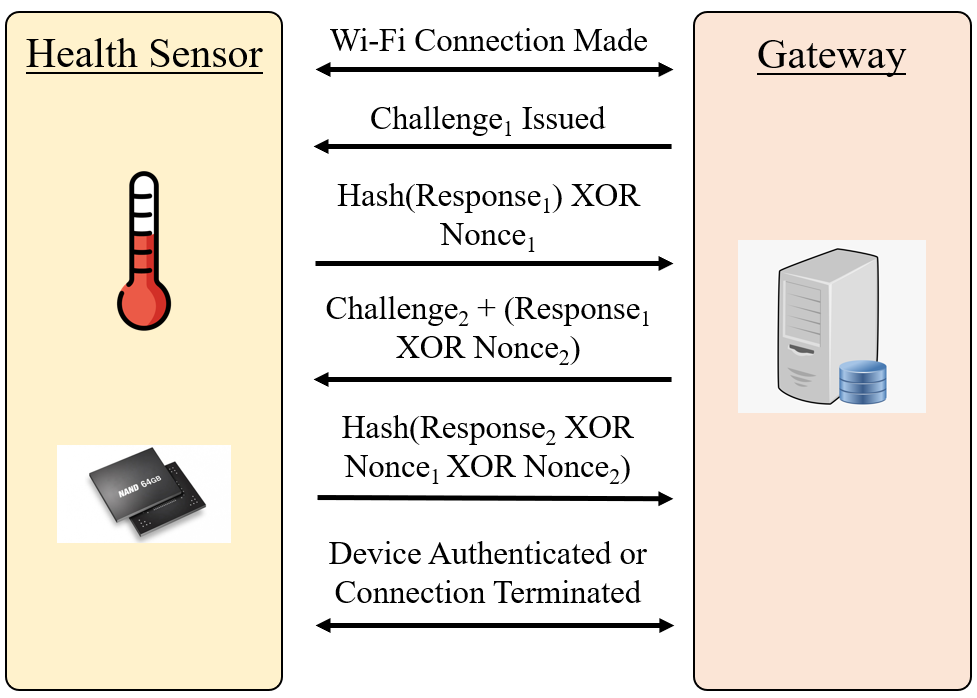}
\caption{Authentication Protocol}
\label{auth-protocol}
\end{figure}

The final step in the HaLo extraction method is authenticating a sensor by issuing a challenge and verifying the response. While most memory based PUF solutions use a challenge-response scheme that links a challenge value with a specific page in memory, HaLo extraction includes additional information in the challenge packet. As mentioned in the \textit{PUF Extraction Observations and Techniques} section, the granularity of the interrupted program is not high enough to generate reliably stable signatures, due to the fact that some cells appear stable in small sample sizes. In order to combat this, the challenge issued to the sensor includes the location in memory, as well as a list of 512 stable byte locations on both the high side and low side of the page programming. It is important to note that these byte indexes do not reveal any polarity of the bytes themselves -- they are simply represent a map of bytes that were identified as extremely stable in the enrollment process. The basic authentication procedure is shown in Figure \ref{auth-protocol}.

\section{Experimental Results and Validations}
In this section, we will discuss the relevant PUF metrics of the novel HaLo extraction process. All of the results generated in this section were gathered at room temperature using the experimental setup discussed in the \textit{Proposed Flash PUF} section. The main characteristics of the HaLo PUF extraction method are its reliability over multiple challenges, the uniqueness of signatures on every page, and the minimal time required in order to generate a signature.

\subsection{PUF Metrics}
\begin{figure}
\center
\includegraphics[width=0.5\linewidth]{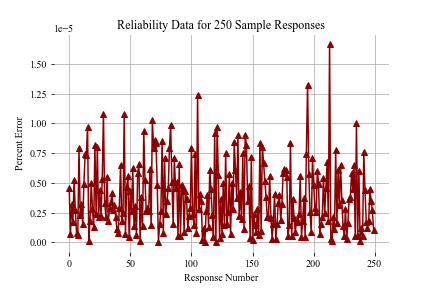}
\caption{Reliability of 250 Responses at $50\%$ of Life Used}
\label{reliability}
\end{figure}

\begin{figure}
\center
\includegraphics[width=0.5\linewidth]{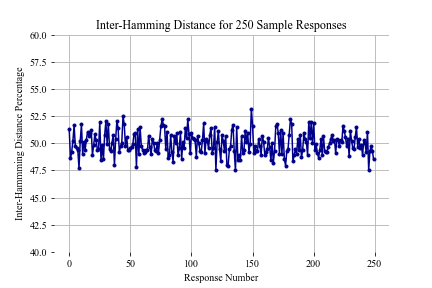}
\caption{Inter-Hamming Distance Calculations between 250 Responses}
\label{inter-hd}
\end{figure}

There are three major metrics for any PUF. They are uniqueness, randomness, and reliability. Uniqueness defines how different each PUF response is from the other. This value is best captured through the Inter-Hamming Distance (Inter-HD). The percentage Inter-HD describes what percentage of bits flip between two different responses. An ideal value for this is approximately $50\%$. Our system had an average of $47\%$ which is shown in Figure~\ref{inter-hd}. The second metric randomness describes how random each signature is. This can be measured in Shannon entropy per bit. With 256 high bits and 256 low bits, the average Shannon entropy per bit is approximately 0.999 with an ideal value of 1. Next, reliability is the final measurement. The reliability of this system is extremely strong. The error does change as the flash cells age; however, the average percent error is approximately $7.1 * 10^{-6}$. The max percent error value is $1.9 * 10^{-2}$ and the minimum value is $5.9 * 10^{-7}$. This reliability is strong enough to possibly support cryptographic key generation and is able to last through the end of the Flash's life this is shown in Figure~\ref{aging} and Figure~\ref{reliability}.


\subsection{Aging Adaptation}
\begin{figure}
\center
\includegraphics[width=0.5\linewidth]{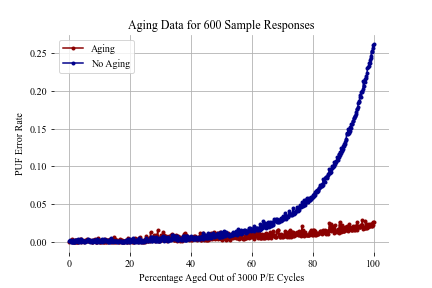}
\caption{Error Rate vs Percentage of Life Used}
\label{aging}
\end{figure}

In order to simulate aging, the Flash memory chips were programmed until their maximum rating which ranges between 3,000 to 4,000 P/E cycles for MLC chips. The aging causes the oxide to deteriorate from the program voltage stress. The low side bytes that are susceptible to over programming are barely modified since these bytes just program and leak charge faster ~\cite{BOCA-NET}. However, the high side bytes that are resistant to programming more easily modify the amount of bit flips in each byte. In general two approaches for this were considered. First was adapting the polling interval and decrease it since the flash cells program faster. However, this approach can be difficult to model and control due to the lack of granularity in the interrupt mechanism. Consequently, the second approach was taken. In this approach, the number of bytes required for a decode was changed once the cells reached a particular percentage of life used. This adapts the error rate and drops it significantly. Instead of taking the majority (or in our case 5 bits) for a decode. The decode threshold was varied according to the life of the flash cells. At $50\%$ of the lifetime, the amount of bits required for a proper decode becomes 6. Then, this changes to a threshold of 7 at $90\%$ which remains fixed and keeps the error rate significantly lower. This trend is reflected in Figure~\ref{aging}. All of the aggregate statistics are reflected in Table~\ref{tabel-stats}.

\subsection{PUF Latency and Power Consumption}
\begin{figure}
\center
\includegraphics[width=0.5\linewidth]{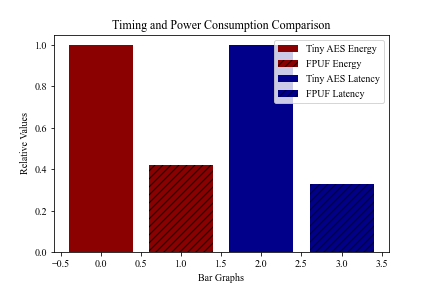}
\caption{Relative Power Consumption and Latency Comparison between the FPUF and Traditional Tiny AES Encryption Scheme}
\label{power}
\end{figure}

 Finally, our PUF will be compared in latency and power consumption to an AES encryption scheme on one of the most portable AES implementations for resource-constrained devices, Tiny AES. To just encrypt a preshared key about 22.3 mW of power is used and the encryption process takes around 190 ms. However, the HaLo extraction technique takes less than 1mW of power and PUF signatures are generated in 34.8 ms. This relative power difference is shown in Figure~\ref{power}. This highlights a significant performance advantage with a higher security guarantee since the PUF signatures are never stored anywhere ~\cite{variations}. Furthermore, this low latency and power consumption make the HaLo PUF a strong candidate for the telehealth application space.
 
\begin{table}
\center
\caption{PUF Entropy, Reliability, and Randomness Metrics}
\label{tab:1}       
\begin{tabular}{llll}
\hline\noalign{\smallskip}
Metric & Minimum & Average & Max  \\
\noalign{\smallskip}\hline\noalign{\smallskip}
Shannon Entropy & 0.99 & 0.99 & 0.99 \\
Reliability & $5.9 * 10^{-7}$ & $7.1 * 10^{-6}$ & $1.9 * 10^{-2}$ \\
 Inter-Hamming Distance& $47.2\%$& $51.0\%$ & $53.1\%$ \\
\noalign{\smallskip}\hline
\label{tabel-stats}
\end{tabular}
\end{table}

\section{Telehealth Application}
The first step in developing our authentication protocol was to build our telehealth application, which involved using a DS18B20 temperature sensor to collect temperature data and store it in a format that we can then use for edge deployment. Body temperature collection is just one of the many different ways remote patient monitoring can be utilized. For our experiment, we used a Raspberry Pi as a means to collect temperature data in intervals of 10 seconds and saved the data in a .csv format. 

We then built a TCP/IP dynamic challenge-response authentication scheme using the PUF. This authentication process starts with the gateway sending a challenge to the health sensor, denoted Challenge$_1$. This request includes a challenge location and stable byte locations of the NAND flash memory cell. The health sensor performs the interrupted program and extracts the stable byte values. The health sensor then randomly generates a nonce, denoted Nonce$_1$, and the generated hash value is XORed with Nonce$_1$. The value generated is sent to the gateway as a response to Challenge$_1$. The gateway knows what Response$_1$ should be, so it XORs the entire response of the gateway with Response$_1$ to determine Nonce$_1$. The gateway then randomly generates another nonce, denoted Nonce$_2$. A second challenge, using a different index of stable byte locations, is generated by the gateway. Challenge$_2$ is then concatenated with (Response$_1$ XOR Nonce$_2$), and this message is sent back to the health sensor. The health sensor is able to separate Challenge$_2$ from (Response$_1$ XOR Nonce$_2$), and (Response$_1$ XOR Nonce$_2$) is XORed with Response$_1$ to determine Nonce$_2$. Both the gateway and the health sensor now know the values of Challenge$_1$, Challenge$_2$, Nonce$_1$, and Nonce$_2$. Using Challenge$_2$, the health sensor again performs the interrupted program and extracts the stable byte values. By the end, if the health sensor and the gateway are legitimate sources, the challenges and responses can be properly decoded and the transaction is authenticated.

There are many advantages to using this authentication scheme. First, the use of hashes masks the plaintext values of the data being sent, meaning that attackers won’t be able to read the data in transit. Second, the use of randomly generated nonces means that the values being transmitted will always change with every transaction. Finally, and most importantly, the use of a PUF provides a unique identifier that the gateway can reliably authenticate, and attackers won’t be able to model the authentication responses of the PUF.

Here we see the challenge and response pairs that we discussed being sent between the health sensor and the gateway. Man-in-the-middle and replay attacks can occur in our any networked environment ~\cite{DVFS, Bio}. However, with regard to MITM, any modification of the packets being sent would result in a complete breakdown of the authentication process, ultimately making it obvious that the connection was tampered with, and would result in the gateway refusing the connection ~\cite{P2M}. With Replay attacks, the use of randomly generated nonces means the values being sent change with every transaction. This means that an attacker can’t replay a message with an old pair of nonces, as the values would be entirely different to the current nonces ~\cite{FLASH, Forte}.

\section{Conclusions and Future Works}
As medical care continues to improve in both effectiveness and accessibility on a global scale, remote patient monitoring is a logical next step for healthcare investment. We have witnessed a global pandemic that disrupted the delivery of potentially life-saving medical care for over a year, which may have been partially mitigated by the adoption of remote patient monitoring (RPM) technology. While the technology proves useful for many medical monitoring applications, it is important to carefully consider the security vulnerabilities and possible exploits of large-scale adoption of RPM. The HaLo extraction method is a lightweight, fast, and easily implementable security measure that could help ensure the authenticity of RPM sensor data. The HaLo method is designed in such a way that it could be implemented on new RPM sensors with minor software updates, and no additional components. The method is resource-efficient and offers financial benefits when compared with many other commercial PUF solutions. 

When considering future work for this project, more extensive testing needs to be done using additional flash chips from different manufacturers. Additionally, testing must be done in high temperature environments in order to verify that the reliability remains high regardless of external factors such as heat. Finally, the HaLo extraction method could be tested on other flash memory densities (SLC/TLC), or possibly even different flash architectures (3-D) in order to broaden the possible application space.


 

\end{document}